\documentclass[10pt,twocolumn,twoside]{article}
\pdfoutput=1
\usepackage{amsmath}
\usepackage{amssymb}

\usepackage{graphicx}

\usepackage{cite}

\usepackage{color} 

\date{}

\usepackage{times}
\usepackage{courier}
\usepackage{underscore}
\usepackage{url}  
\usepackage{placeins}  
\usepackage{setspace}
\usepackage[labelfont=bf,labelsep=period,skip=3pt,font={footnotesize,stretch=0.99}]{caption}
\usepackage{hyperref}
\usepackage{fancyhdr}
\newcommand{\R}{{\textit{R}}}
\newcommand{\code}[1]{{\texttt{#1}}}

\begin{document}
\title{Waste Not, Want Not: \\ Why Rarefying Microbiome Data Is Inadmissible}
\maketitle

\vskip0.5cm
Paul J. McMurdie$^{1}$, 
Susan Holmes$^{1*}$
\\
${ }^\mathbf{1}$ Statistics Department, Stanford University\\
Stanford, CA, USA
\\
$\ast$ E-mail: susan@stat.stanford.edu

\begin{abstract}
Current practice in the normalization of
microbiome count data is inefficient in the statistical sense.
For apparently historical reasons, 
the common approach is either to use simple proportions
(which does not address heteroscedasticity)
or to use \emph{rarefaction} of  counts,
even though both of these approaches are inappropriate
for detection of differentially abundant species.
Well-established statistical theory is available
that simultaneously accounts for library size differences and biological variability
using an appropriate mixture model.
Moreover, specific implementations for DNA sequencing read count data 
(based on a Negative Binomial model for instance)
are already available in RNA-Seq focused R packages
such as edgeR and DESeq.
Here we summarize the supporting statistical theory, 
and use simulations and empirical data
to precisely demonstrate the substantial improvements 
provided by a relevant mixture model approach
over simple proportions or rarefying.
We show how both proportions and rarefied counts
result in a high rate of false positives
in tests for species that are differentially abundant across sample classes.
Regarding microbiome sample-wise clustering,
we also show that the rarefying procedure
often discards samples that can be accurately clustered 
by alternative methods.
We further compare different Negative Binomial methods
with a recently-described zero-inflated Gaussian mixture, 
implemented in a package called \emph{metagenomeSeq}.
We find that metagenomeSeq performs well 
when there is an adequate number of biological replicates,
but nevertheless tends toward a higher false positive rate,
a key trade-off that may be critical for the goals of different investigations. 
Based on these results and well-established statistical theory, 
we advocate that investigators avoid rarefying altogether.
We have provided microbiome-specific extensions 
to these tools in the R package, phyloseq.
\end{abstract}

\section*{Author Summary}
%
The term \emph{microbiome} refers to the ecosystem of microbes
that live in a defined environment.
The decreasing cost and increasing speed of DNA sequencing technology
has recently provided scientists with affordable and timely access
to the genes and genomes of microbiomes 
that inhabit our planet and even our own bodies.
In these investigations many microbiome samples
are sequenced at the same time on the same DNA sequencing machine,
but often result in total numbers of sequences per sample that are vastly different.
The common procedure for addressing this difference in sequencing effort across samples --
different \emph{library sizes} --
is to either (1) base analyses on the proportional abundance of each species in a library,
or  (2) \emph{rarefy}, throw away sequences from the larger libraries 
so that all have the same, smallest size.
We show that both of these \emph{normalization methods} 
sometimes work acceptably well for the purpose of 
comparing entire microbiomes to one another,
but that neither method works well when comparing 
the relative proportions of each bacterial species
across microbiome samples.
We show that alternative methods
based on a statistical \emph{mixture model} perform very well,
and can be easily adapted from a separate biological sub-discipline, called \mbox{RNA-Seq} analysis.



\section*{Introduction}

Modern, massively parallel DNA sequencing technologies
have changed the scope and technique of investigations 
across many fields of biology~\cite{Shendure:2012et,Shendure:2008jh}. 
In gene expression studies the standard measurement technique 
has shifted away from microarray hybridization 
to direct sequencing of cDNA,
a technique often referred to as \emph{RNA-Seq}~\cite{RNASeq}. 
Analogously, culture independent~\cite{Pace:1997tl} microbiome research
has migrated away from detection of species
through microarray hybridization of 
small subunit rRNA gene PCR amplicons~\cite{PhyloChip}
to direct sequencing of highly-variable regions of these amplicons~\cite{Huse:2008kw},
or even direct \emph{shotgun} sequencing of microbiome metagenomic DNA~\cite{shotgun}.
Even though the statistical methods
available for analyzing microarray data
have matured to a high level of sophistication~\cite{Allison:2006fk},
these methods are not directly applicable
because DNA sequencing data consists of discrete counts
of equivalent sequence \emph{reads}
rather than continuous values derived from
the fluorescence intensity of hybridized probes.
In recent generation DNA sequencing the total reads per sample 
(\emph{library size}; sometimes referred to as \emph{depths of coverage})
can vary by orders of magnitude within a single sequencing run. 
Comparison across samples with different library sizes requires
more than a simple linear or logarithmic scaling adjustment
because it also implies different levels of uncertainty, 
as measured by the sampling variance 
of the proportion estimate for each feature 
(a feature is a gene in the RNA-Seq context,
and is a species or \underline{O}perational \underline{T}axonomic \underline{U}nit, OTU,
in the context of microbiome sequencing).
In this article we are primarily concerned with 
optimal methods for addressing differences in library sizes 
from microbiome sequencing data.

Variation in the read counts of features
between technical replicates 
have been adequately modeled by Poisson random variables~\cite{Marioni2008}.
However, we are usually interested in understanding
the variation of features among biological replicates
in order to make inferences that are relevant to the corresponding population;
in which case a mixture model is necessary
to account for the added uncertainty~\cite{Lu:2005ea}.
Taking a hierarchical model approach with the Gamma-Poisson
has provided a satisfactory fit to RNA-Seq data~\cite{Robinson:2007fv},
as well as a valid regression framework that leverages
the power of generalized linear models~\cite{Cameron:2013tp}.
A Gamma mixture of Poisson variables gives
the negative binomial (NB) distribution~\cite{Lu:2005ea,Robinson:2007fv} and 
several RNA-Seq analysis packages now model the counts, $K$,
for gene $i$, in sample $j$ according to:

\begin{equation}
K_{ij} \sim \operatorname{NB}(s_{j}\mu_{i}, \phi_{i})
\end{equation}

where
$s_{j}$ is a linear scaling factor for sample $j$ that accounts for its library size,
$\mu_{i}$ is the mean proportion for gene $i$, and
$\phi_{i}$ is the dispersion parameter for gene $i$.
The variance is $\nu_{i} = s_j \mu_{i} + \phi_{i}s_j^2 \mu_{i}^{2}$,
with the NB distribution becoming Poisson when $\phi = 0$.
Recognizing that $\phi > 0$ and estimating its value
is important in gene-level tests,
in order to better control the rate of false positive genes 
that test as significantly differentially expressed
between experimental conditions under the assumption of a Poisson distribution, 
but nevertheless fail in tests that account for non-zero dispersion.

The uncertainty in estimating $\phi_{i}$ for every gene
when there is a small number of samples ---
or a small number of biological replicates ---
can be mitigated by sharing information 
across the thousands of genes in an experiment,
leveraging a systematic trend in the mean-dispersion relationship~\cite{Robinson:2007fv}.
This approach substantially increases the power 
to detect differences in proportions (differential expression)
while still adequately controlling for false positives~\cite{DESeq}. 
Many R packages implementing this model of RNA-Seq data are now available,
differing mainly in their approach to modeling dispersion across genes~\cite{sSeq}.
Although DNA sequencing-based microbiome investigations
use the same sequencing machines
and represent the processed sequence data in the same manner --- 
a feature-by-sample contingency table 
where the features are OTUs instead of genes --- 
to our knowledge the modeling and normalization methods
currently used in RNA-Seq analysis 
have not been transferred to microbiome research~\cite{DiBella:2013ic,Segata:2013cg,NavasMolina:2013bc}.

Instead, microbiome analysis workflows often begin
with an \emph{ad hoc} library size normalization
by random subsampling without replacement,
or so-called \emph{rarefying}~\cite{Hughes:2005hq,Koren:2013dv,NavasMolina:2013bc}. 
There is confusion in the literature regarding terminology,
and sometimes this normalization approach 
is conflated with a non-parametric resampling technique --- 
called \emph{rarefaction}~\cite{Sanders:1968},
or \emph{individual-based taxon re-sampling curves}~\cite{Gotelli:2001ho} ---
that can be justified for coverage analysis 
or species richness estimation in some settings~\cite{Gotelli:2001ho},
though in other settings it can perform worse than parametric methods~\cite{Mao:2005cl}.
Here we emphasize the distinction between 
taxon re-sampling curves and normalization
by strictly adhering to the terms \emph{rarefying} or \emph{rarefied counts}
when referring to the normalization procedure,
and respecting the original definition for \emph{rarefaction}.
Rarefying is most often defined by the following steps~\cite{Hughes:2005hq}.
\begin{enumerate}
\item Select a minimum library size, $N_{L,min}$.
This has been called the \emph{rarefaction level}~\cite{NavasMolina:2013bc},
though we will not use the term here.
\item Discard libraries (samples) that have fewer reads than $N_{L,min}$.
\item Subsample the remaining libraries without replacement
such that they all have size $N_{L,min}$. 
\end{enumerate}
Often $N_{L,min}$ is chosen to be equal 
to the size of the smallest library
that is not considered \emph{defective},
and the process of identifying defective samples
comes with a risk of subjectivity and bias.
In many cases researchers have also failed to repeat the random subsampling step
or record the pseudorandom number generation seed/process --- 
both of which are essential for reproducibility.
To our knowledge, rarefying was first recommended for microbiome counts
in order to moderate the sensitivity of the UniFrac distance~\cite{Lozupone:2005unifrac}
to library size, 
especially differences in the presence of rare OTUs~\cite{Lozupone:2010dh}.
In these and similar studies the principal objective
is an exploratory/descriptive comparison of microbiome samples,
often from different environmental/biological sources;
a research task that is becoming increasingly accessible
with declining sequencing costs
and the ability to sequence many samples in parallel
using barcoded primers~\cite{Hamady:2008iu,Liu:2008bi}.
Rarefying is now an exceedingly common precursor
to microbiome multivariate workflows
that seek to relate sample covariates 
to sample-wise distance matrices~\cite{Hamady:2009,Yatsunenko:2012gi,Koren:2013dv};
for example, integrated as a recommended option in QIIME's~\cite{caporaso2010qiime}
\texttt{beta\_diversity\_through\_plots.py} workflow,
in \texttt{Sub.sample} in the mothur software library~\cite{mothur},
in \texttt{daisychopper.pl}~\cite{Gilbert:2009kv},
and is even supported in phyloseq's \texttt{rarefy\_even\_depth} function~\cite{phyloseqplosone}
(though not recommended in its documentation).
The perception in the microbiome literature of ``rarefying to even sampling depth''
as a standard normalization procedure
appears to explain why \emph{rarefied} counts are also used
in studies that attempt to detect \emph{differential abundance} of OTUs
between predefined classes of samples~\cite{Charlson:2010ij,Price:2010hw,building:microbes,Flores:2013dq,Kang:2013gj},
in addition to studies that use proportions directly~\cite{Segata:2012cm}.
It should be noted that we have adopted 
the recently coined term \emph{differential abundance}~\cite{metastats,metagenomeSeq}
as a direct analogy to \emph{differential expression} from RNA-Seq. 
Like differentially expressed genes, 
a species/OTU is considered differentially abundant 
if its mean proportion is significantly different
between two or more sample classes
in the experimental design.

%

\subsection*{Statistical motivation}
Despite its current popularity in microbiome analyses
\textbf{rarefying biological count data is statistically inadmissible}
because it requires the omission of available valid data.
This holds even if repeated rarefying trials are compared 
for stability as previously suggested~\cite{NavasMolina:2013bc}.
In this article we demonstrate the applicability of
a variance stabilization technique 
based on a mixture model of microbiome count data.
This approach simultaneously addresses both problems of 
(1) DNA sequencing libraries of widely different sizes, and 
(2) OTU (feature) count proportions that vary more than expected under a Poisson model. 
We utilize the most popular implementations
of this approach currently used in RNA-Seq analysis,
namely edgeR~\cite{edgeR} and DESeq~\cite{DESeq}, 
adapted here for microbiome data.
This approach allows valid comparison across OTUs
while substantially improving both power and accuracy
in the detection of differential abundance.
We also compare the Gamma-Poisson mixture model performance
against a method that models OTU proportions 
using a zero-inflated Gaussian distribution,
implemented in a recently-released package called metagenomeSeq~\cite{metagenomeSeq}.

\begin{table}[!ht]
\centering
\captionsetup{skip=1pt}
\includegraphics[width=3in]{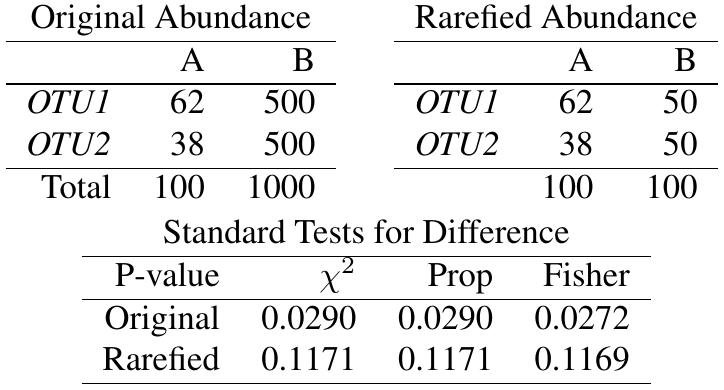}
\caption{
\textbf{A minimal example of the effect of rarefying on power.}
Hypothetical abundance data in its original (Top-Left) and rarefied (Top-Right) form,
with corresponding formal test results for differentiation (Bottom).
}
\label{tab:rarefyminimal}
\end{table}

A mathematical proof of the sub-optimality of the
subsampling approach is presented in the supplementary material
(Text~S\ref{supp:mathstat}).
To help explain why rarefying is statistically inadmissible,
especially with regards to variance stabilization,
we start with the following minimal example.
Suppose we want to compare two different samples,
called \emph{A} and \emph{B},
comprised of 100 and 1000 DNA reads, respectively.
In statistical terms, these library sizes are also equivalent to the 
number of trials in a sampling experiment.
In practice, the library size associated with each biological sample
is a random number generated by the technology,
often varying from hundreds to millions.
For our example, we imagine the simplest possible case
where the samples can only contain two types of microbes,
called \emph{OTU1} and \emph{OTU2}.
The results of this hypothetical experiment are represented
in the \emph{Original Abundance} section of Table~\ref{tab:rarefyminimal}.
Formally comparing the two proportions according to a standard test
could technically be done either using a $\chi^2$ test
(equivalent to a two sample proportion test here) 
or a Fisher exact test.
By first rarefying (Table~\ref{tab:rarefyminimal}, \emph{Rarefied Abundance} section) 
so that both samples have the same library size before doing the tests,
we are no longer able to differentiate the samples 
(Table~\ref{tab:rarefyminimal}, tests).
This loss of power is completely attributable
to reducing the size of \emph{B} by a factor of 10,
which also increases the confidence intervals 
corresponding to each proportion
such that they are no longer distinguishable
from those in \emph{A},
even though they are distinguishable in the original data.

The variance of the proportion's estimate $\hat{p}$
is multiplied by 10 when the total count 
is divided by 10. 
In this binomial example 
the variance of the proportion estimate is
$Var(\frac{X}{n})=\frac{pq}{n}=\frac{q}{n}E(\frac{X}{n})$,
a function of the mean. 
This is a common occurrence and one that
is traditionally dealt with in statistics by applying
variance-stabilizing transformations.
We show in Text~S\ref{supp:mathstat}
that the relation between the variance and the mean
for microbiome count data can be estimated
and the model used to find the optimal variance-stabilizing transformation.
As illustrated by this simple example,
it is inappropriate to compare
the proportions of OTU $i$, $p_i=K_{ij}/s_{j}$, 
without accounting for differences in the denominator value
(the library size, $s_{j}$)
because they have unequal variances.
This problem of unequal variances is called \emph{heteroscedasticity}.
In other words, the \emph{uncertainty} associated
with each value in the table
is fundamentally linked to the total number of observations (or reads),
which can vary even more widely than a 10-fold difference.
In practice we will be observing 
hundreds of different OTUs instead of two,
often with dependendency between the counts.
Nevertheless, the difficulty caused by unequal library 
sizes still pertains.

The uncertainty with which each proportion is estimated must be considered when testing
for a difference between proportions (one OTU),
or sets of proportions (a microbial community).
Although rarefying does equalize variances,
it does so only by inflating the variances in all samples
to the largest (worst) value among them
at the cost of discriminating power (increased uncertainty).
Rarefying adds additional uncertainty
through the random subsampling step,
such that Table~\ref{tab:rarefyminimal} shows the best-case, 
approached only with a sufficient number of \emph{repeated} rarefying trials
(See Protocol~S\ref{supp:rmdzip}, minimal example).
In this sense alone, the random step in rarefying is unnecessary.
Each count value could be transformed to a \emph{common-scale}
by rounding $K_{ij}s_{min}/s_{j}$.
Although this common-scale approach is an improvement
over the rarefying method here defined,
both methods suffer from the same problems related to lost data.

\begin{figure}[!htbp]
\begin{flushleft}
\includegraphics[width=3in]{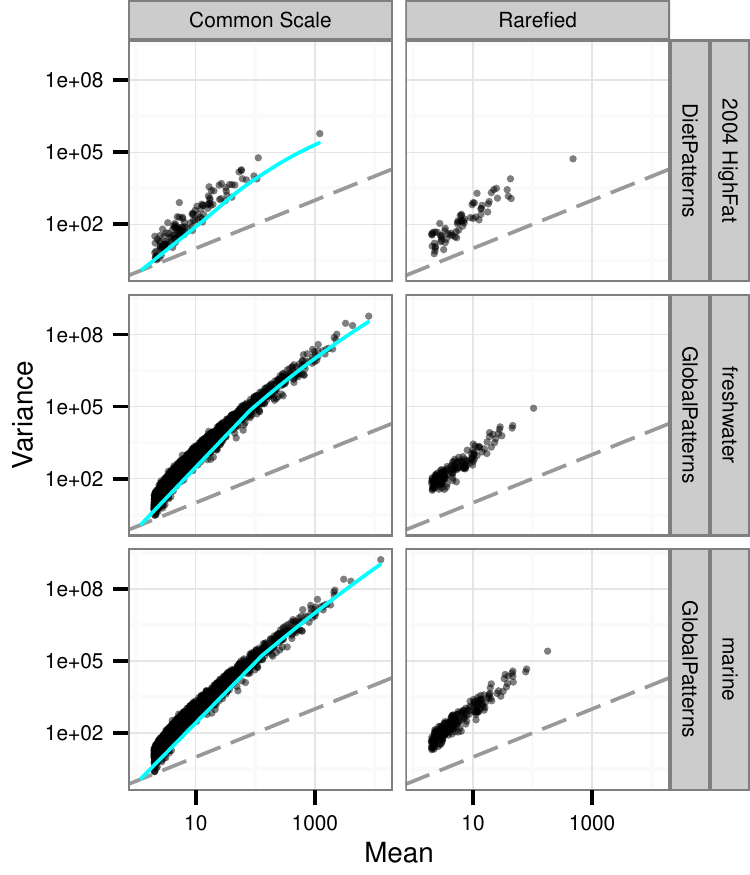}
\end{flushleft}
\caption{
\textbf{Overdispersion in Microbiome Data.}
Common-Scale Variance versus Mean for Microbiome Data.
Each point in each panel represents 
a different OTU's mean/variance estimate
for a biological replicate and study. 
The data in this figure come from 
the \emph{Global Patterns} survey~\cite{global patterns}
and the \emph{Long-Term Dietary Patterns} study~\cite{DietPatterns},
with results from many more studies included in 
Protocol~S\ref{supp:rmdzip}.
(Right) Variance versus mean abundance for rarefied counts. 
(Left) Common-scale variances and common-scale means,
estimated according to Equations 7 and 6 from Anders and Huber~\cite{DESeq},
implemented in the DESeq package (Text~S\ref{supp:mathstat}).
The dashed gray line denotes the $\sigma^2 = \mu$ case
(Poisson; $\phi = 0$).
The cyan curve denotes the fitted variance estimate using DESeq~\cite{DESeq},
with \texttt{method=`pooled', sharingMode=`fit-only', fitType=`local'}. 
}
\label{fig:overdispersion}
\end{figure}

\clearpage
\section*{Materials and Methods}

\begin{figure*}[!htbp]
\centering
\includegraphics[width=4.5in]{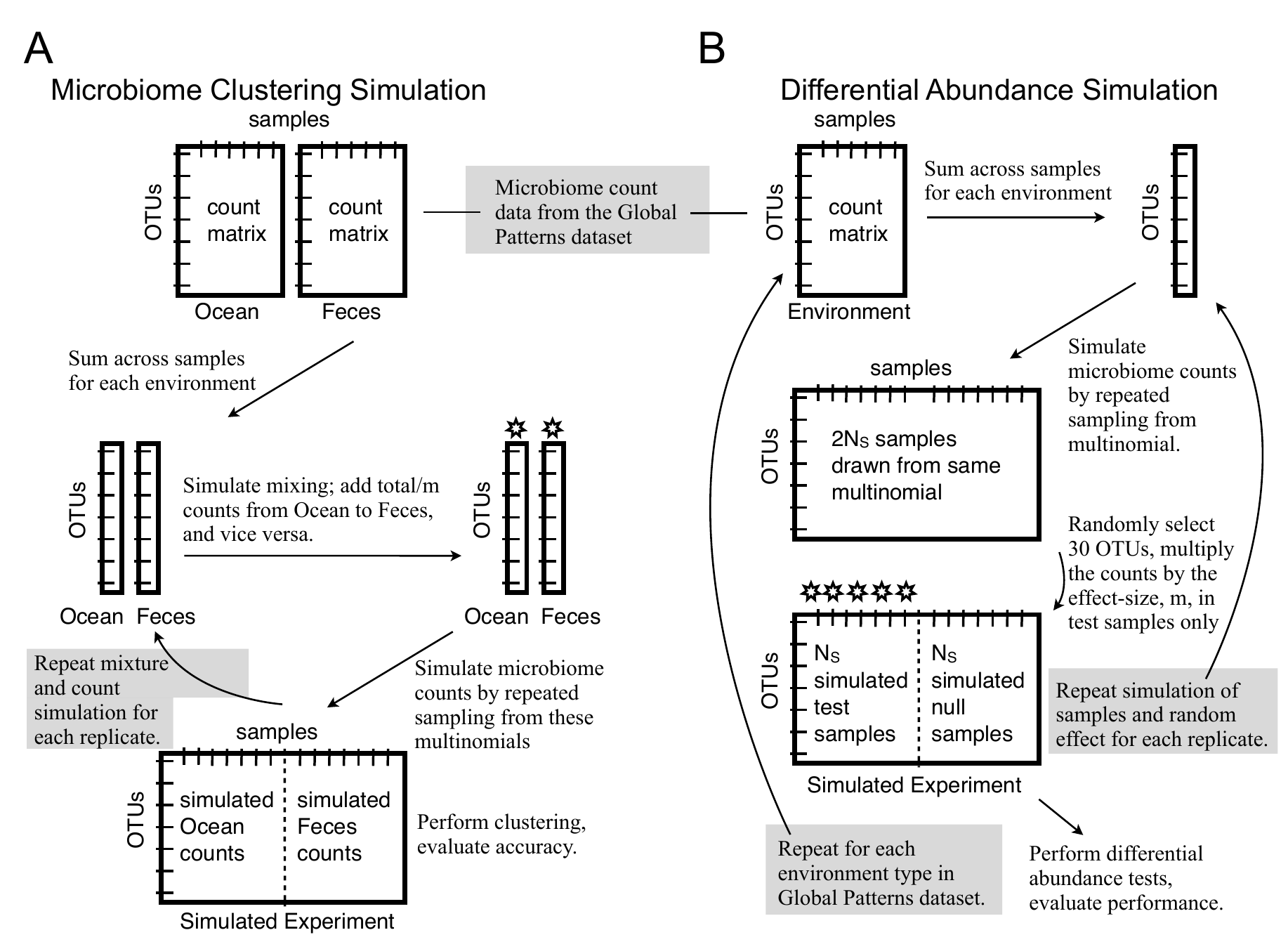}
\caption{
{\bf Graphical summary of the two simulation frameworks.}
Both \emph{Simulation A} (clustering)
and \emph{Simulation B} (differential abundance) 
are represented. 
All simulations begin with real microbiome count data 
from a survey experiment
referred to here as ``the Global Patterns dataset''~\cite{globalpatterns}.
A rectangle with tick marks and index labels (top or left)
represents an abundance count matrix (``OTU table''),
while a much thinner rectangle with only OTU tick marks
represents a multinomial of OTU counts/proportions.
In both simulation designs,
the variable \emph{m} is used to refer to the effect size,
but its meaning is different in each simulation.
Small stars emphasize a multinomial or sample
in which a perturbation (our effect) has been applied.
}
\label{fig:simdesign}
\end{figure*}

In order to quantify the relative statistical costs of rarefying,
and to illustrate the relative benefits of an appropriate mixture model,
we created two microbiome simulation workflows 
based on repeated subsampling from empirical data.
These workflows were organized according to Figure~\ref{fig:simdesign}.
Because the correct answer in every simulation is known,
we were able to evaluate the resulting
power and accuracy of each statistical method,
and thus quantify the improvements one method provided over another
under a given set of conditions.
In both simulation types we varied the library and effect sizes
across a range of levels 
that are relevant for recently-published microbiome investigations,
and followed with commonly used statistical analyses
from the microbiome and/or RNA-Seq literature
(Figure~\ref{fig:simdesign}).

\subsection*{Simulation A}
Simulation~A is a simple example of a descriptive experiment
in which the main goal is to distinguish patterns of relationships
between whole microbiome samples
through normalization followed by the
calculation of sample-wise distances. 
Many early microbiome investigations are variants of Simulation~A,
and also used rarefying prior to calculating UniFrac distances~\cite{Hamady:2009}.
Microbiome studies have often graphically represented 
the results of their pairwise sample distances
using multidimensional scaling~\cite{gower:1966hi}
(also called \emph{Principal Coordinate Analysis}, PCoA),
which is useful if the desired effects
are clearly evident among the first two or three ordination axes.
In some cases, formal testing of sample covariates 
is also done using a permutation MANOVA
(e.g. \texttt{vegan::adonis} in R~\cite{veganpkg})
with the (squared) distances and covariates 
as response and linear predictors, respectively~\cite{anderson2001new}.
However, in this case we are not interested 
in creating summary graphics or testing the explanatory power of sample covariates,
but rather we are interested in precisely evaluating 
the relative discriminating capability 
of each combination of normalization method and distance measure.
We will use clustering results as a quantitative proxy
for the broad spectrum of approaches taken to interpret
microbiome sample distances.

\textbf{Normalizations in Simulation A}.
For each simulated experiment we used the following normalization methods prior to calculating sample-wise distances.
\begin{enumerate}
\item \textbf{DESeqVS}. Variance Stabilization implemented in the DESeq package~\cite{DESeq}. 
\item \textbf{None}. Counts not transformed. Differences in total library size could affect the values of some distance metrics.
\item \textbf{Proportion}. Counts are divided by total library size.
\item \textbf{Rarefy}. Rarefying is performed as defined in the introduction, using \texttt{rarefy\_even\_depth} implemented in the phyloseq package~\cite{phyloseqplosone}, with $N_{L,min}$ set to the 15$^{th}$-percentile of library sizes within each simulated experiment.
\item \textbf{UQ-logFC}. The \emph{Upper-Quartile Log-Fold Change} normalization implemented in the edgeR package~\cite{edgeR}, coupled with the \emph{top-MSD} distance (see below).
\end{enumerate}

\textbf{Distances in Simulation A}. 
For each of the previous normalizations we calculated sample-wise distance matrices using the following distance metrics, if applicable.
\begin{enumerate}
\item \textbf{Bray-Curtis}. The Bray-Curtis distance first defined in 1957 for forest ecology~\cite{Bray:1957hk}. 
\item \textbf{Euclidean}. The euclidean distance treating each OTU as a dimension.
This has the form $\sqrt{\sum_{i=1}^{n}(K_{i1}-K_{i2})^2}$, 
for the distance between samples $1$ and $2$, 
with $K$ and $i$ as defined in the Introduction
and $n$ the number of distinct OTUs.
\item \textbf{PoissonDist}. Our abbreviation of \texttt{PoissonDistance}, a sample-wise distance implemented in the PoiClaClu package~\cite{Witten2011}.
\item \textbf{top-MSD}. The mean squared difference of top OTUs, as implemented in edgeR~\cite{edgeR}.
\item \textbf{UniFrac-u}. The Unweighted UniFrac distance~\cite{Lozupone:2005unifrac}.
\item \textbf{UniFrac-w}. The Weighted UniFrac distance~\cite{Lozupone:2007kg}.
\end{enumerate}

In order to consistently evaluate performance in this regard,
we created a simulation framework in which there are only two templates
and each microbiome sample is drawn from one of these templates by sampling with replacement.
The templates originate from the \emph{Ocean} and \emph{Feces} samples
of the \emph{Global Patterns} empirical dataset~\cite{globalpatterns}.
These two datasets were chosen because they have negligible overlapping OTUs,
allowing us to modify the severity of the difference between the templates
by randomly mixing a proportion of counts between the Ocean and Feces data
prior to generating a set of samples for each simulated experiment.
This mixing step allows arbitrary control over the difficulty 
of the sample classification task 
from trivial (no mixing) to impossible (evenly mixed).
Unsupervised classification was performed independently 
for each combination of simulated experiment, normalization method, and distance measure
using partitioning around medoids (PAM~\cite{Kaufmancluster,clustercomp},
an alternative to k-means that is considered more robust)
with the number of classes fixed at two.
The accuracy in the classification results
was defined as the fraction of simulated samples correctly classified;
with the worst possible accuracy being 50\% if all samples are given a classification.
Note that the rarefying procedure omits samples,
so its accuracy can be below 50\% under this definition.

The number of samples to include for each template in Simulation~A
was chosen arbitrarily after some exploration of preliminary simulations.
It was apparent that the classification results from Simulation~A were most informative 
when we included enough samples per simulated experiment to achieve reproducible results,
but not so many that it was experimentally unrealistic and prohibitively slow to compute.
Conversely, the preliminary classification results from Simulation~A
that included only a few samples per experiment 
presented a large variance on each performance measure that was difficult to interpret.

\subsection*{Simulation B}
Simulation~B is a simple example of microbiome experiments 
in which the goal is to detect microbes that are
differentially abundant between two pre-determined classes of samples.
This experimental design appears in many clinical settings
(health/disease, target/control, etc.),
and other settings for which there is sufficient 
\emph{a priori} knowledge about the microbiological conditions,
and we want to enumerate the OTUs that are different between these microbiomes,
along with a measure of confidence that the proportions differ.
For this class of analysis, we simulated microbiome samples
by sampling with replacement from a single empirical source environment 
in the Global Patterns dataset.
The samples were divided into two equally-sized classes, target and control,
and a perturbation was applied (multiplication by a defined value)
to the count values of a random subset of OTUs in the target samples.
Each of the randomly perturbed OTUs is differentially abundant between the classes, 
and the performance of downstream tests can be evaluated 
on how well these OTUs are detected without falsely selecting OTUs
for which no perturbation occurred (false positives).
This approach for generating simulated experiments with a defined effect size
(in the form of multiplicative factor) 
was repeated for each combination of 
median library size, number of samples per class,
and the nine available source environments in the Global Patterns dataset.
Each simulated experiment was subjected to various approaches 
for normalization/noise-modeling and differential abundance testing.
False negatives are perturbed OTUs that went undetected,
while false positives are OTUs that were labeled significantly differentially abundant by a test,
but were actually unperturbed and therefore had the same expected proportion in both classes.

\textbf{Normalization/Modeling in Simulation B}. 
For each simulated experiment,
we used the following normalization/modeling methods 
prior to testing for differential abundance.
\begin{enumerate}
\item \textbf{Model/None}. A parametric model was applied to the data, or,
in the case of the t-test, no normalization was applied
(note: the t-test without normalization can only work with 
a high degree of balance between classes,
and is provided here for comparison but is not recommended in general).
\item \textbf{Rarefied}. Rarefying is performed as defined in the introduction, using \texttt{rarefy\_even\_depth} implemented in the phyloseq package~\cite{phyloseqplosone}, with $N_{L,min}$ set to the 15$^{th}$-percentile of library sizes within each simulated experiment.
\item \textbf{Proportion}. Counts are divided by total library size.
\end{enumerate}

\textbf{Testing in Simulation B}.
For each OTU of each simulated experiment 
we used the following to test for differential abundance. 
\begin{enumerate}
\item \textbf{two sided Welch t-test}. A two-sided t-test with unequal variances,
using the \texttt{mt} wrapper in phyloseq~\cite{phyloseqplosone}
of the \texttt{mt.maxT} method in the multtest package~\cite{multtest}.
\item \textbf{edgeR - exactTest}. An exact binomial test (see base R's \texttt{stats::binom.test}) 
generalized for overdispersed counts~\cite{Robinson:2007fv} 
and implemented in the \texttt{exactTest} method of the edgeR package~\cite{edgeR}.
\item \textbf{DESeq - nbinomTest}. A Negative Binomial conditioned test similar to the edgeR test above,
implemented in the \texttt{nbinomTest} method of the DESeq package~\cite{DESeq}.
See the subsection \emph{Testing for differential expression} 
in Anders and Huber, 2010~\cite{DESeq} for the precise definition.
\item \textbf{DESeq2 - nbinomWaldTest}. A Negative Binomial Wald Test
using standard maximum likelihood estimates for GLM coefficients
assuming a zero-mean normal prior distribution,
implemented in the \texttt{nbinomWaldTest} method of the DESeq2 package.
\item \textbf{metagenomeSeq - fitZig}. An Expectation-Maximization estimate
of the posterior probabilities of differential abundance
based on a \emph{Zero Inflated Gaussian} model,
implemented in the \texttt{fitZig} method of the metagenomeSeq package~\cite{metagenomeSeq}. 
\end{enumerate}
All tests were corrected for multiple inferences using the Benjamini-Hochberg method
to control the False Discovery Rate~\cite{BenjaminiHochberg}.
It should be noted that the library sizes 
for both categories of simulation
were sampled from the original distribution of library sizes in the Global Patterns dataset, 
and then scaled according to the prescribed median library size of each simulated experiment.

We have included in Protocol~S\ref{supp:rmdzip}
the complete source code for computing
the survey, simulations, normalizations, and performance assessments
described in this article. 
Where applicable, this code includes the RNG seed so that the simulations
and random resampling procedures
can be reproduced \emph{exactly}.
Interested investigators can inspect and modify this code, 
change the random seed and other parameters,
and observe the results (including figures).
For ease of inspection, we have authored the source code
in \emph{R flavored markdown}~\cite{Rmarkdown},
through which we have generated HTML5 files for each simulation
that include our extensive comments interleaved 
with code, results, and both intermediate and final figures.
Our simulation output can be optionally-modified and re-executed
using the the \texttt{knit2html} function in the knitr package.
This function will take the location of the simulation source files as input,
evaluate its R code in sequence, generate graphics and markdown,
and produce the complete HTML5 output file 
that can be viewed in any modern web browser.
These simulations, analyses, and graphics rely upon the 
cluster~\cite{clusterRpkg},
foreach~\cite{foreach},
ggplot2~\cite{ggplot2}, 
metagenomeSeq~\cite{metagenomeSeq},
phyloseq~\cite{phyloseqplosone}, 
plyr~\cite{plyr},
reshape2~\cite{reshape},
and 
ROCR~\cite{ROCR}
R packages; 
in addition to the 
DESeq(2)~\cite{DESeq},
edgeR~\cite{edgeR},
and  
PoiClaClu~\cite{Witten2011}
R packages for RNA-Seq data, 
and tools available in the standard R distribution~\cite{Rlanguage}.
The Global Patterns~\cite{globalpatterns} dataset
included in phyloseq was used as empirical microbiome template data.
The code to perform the survey and generate Figure~\ref{fig:overdispersion}
is also included as a R Markdown source file in
Protocol~S\ref{supp:rmdzip},
and includes the code to acquire the data
using the phyloseq interface 
to the \url{microbio.me/qiime} server,
a function called \code{microbio\_me\_qiime}.

\begin{figure*}[!htbp]
\begin{center}
\includegraphics[width=6in]{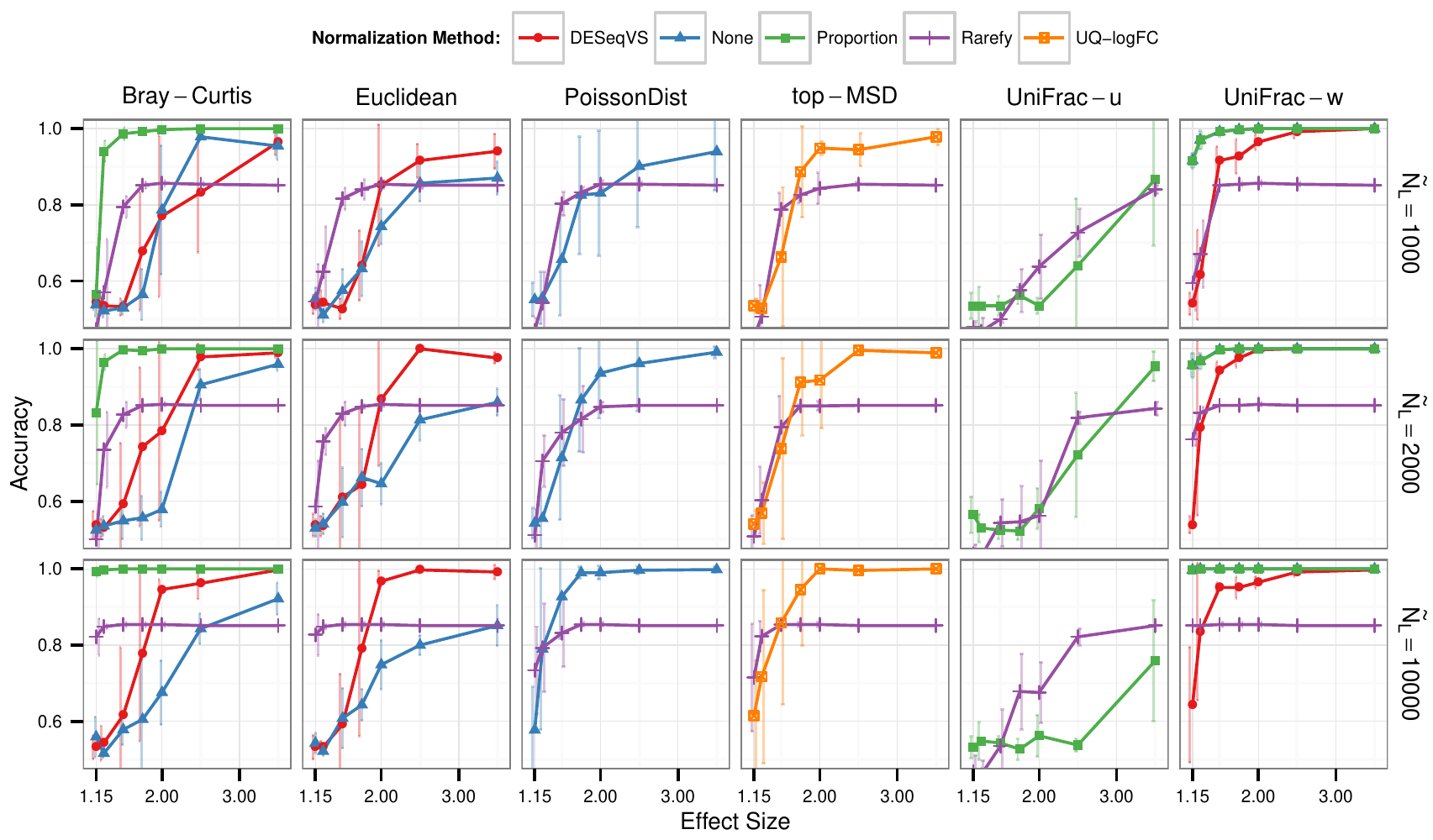}
\end{center}
\caption{
{\bf Clustering accuracy in simulated two-class mixing.} 
Partitioning around medoids (PAM)~\cite{Kaufmancluster,clustercomp}
clustering accuracy (vertical axis) that results
following different normalization and distance methods.
Points denote the mean values of replicates,
with a vertical bar representing one standard deviation above and below.
Normalization method is indicated by both shade and shape,
while panel columns and panel rows indicate
the distance metric and median library size ($\tilde{N}_L$), respectively.
The horizontal axis is the effect size,
which in this context is an \emph{unmixed factor},
the ratio of target to non-target simulated counts
between two microbiomes that effectively have no overlapping OTUs
(Fecal and Ocean microbiomes in the Global Patterns dataset~\cite{globalpatterns}).
Higher values of effect size indicate an easier clustering task.
For precise definitions of abbreviations see
\emph{Simulation~A} of the \emph{Materials and Methods} section.
}
\label{fig:clustaccuracy}
\end{figure*}

\clearpage
\section*{Results and Discussion}

\begin{figure*}[!htbp]
\begin{center}
\includegraphics[width=5.6in]{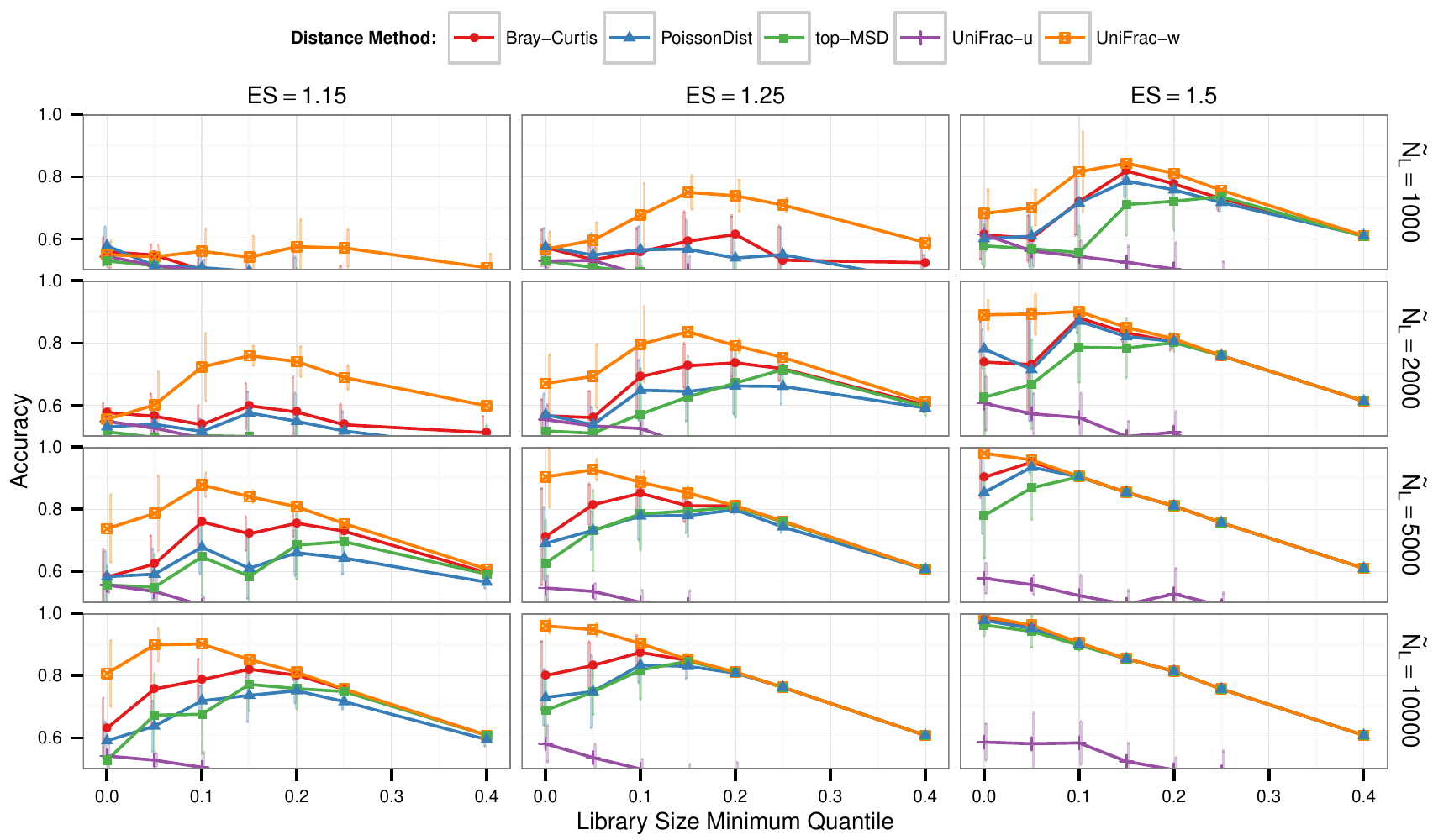}
\end{center}
\caption{
{\bf Normalization by rarefying only, dependency on library size threshold.}
Unlike the analytical methods represented in Figure~\ref{fig:clustaccuracy},
here rarefying is the only normalization method used,
but at varying values of the minimum library size threshold,
shown as library-size quantile (horizontal axis). 
Panel columns, panel rows, and point/line shading indicate 
effect size (ES), 
median library size ($\tilde{N}_L$),
and distance method applied after rarefying, respectively.
Because discarded samples cannot be accurately clustered,
the line $y=1-x$ represents the maximum achievable accuracy.
}
\label{fig:clustaccuracyrare}
\end{figure*}

We performed a survey of publicly available microbiome count data,
to evaluate the variance-mean relationship 
for OTUs among sets of biological replicates
(Figure~\ref{fig:overdispersion}).
In every instance the variances were larger 
than could be expected under a Poisson model
(overdispersed, $\phi > 0$),
especially at larger values of the common-scale mean.
By definition, these OTUs are the most abundant,
and receive the greatest interest in many studies.
For rarefied counts the absolute scales are decreased 
and there are many fewer OTUs that pass filtering,
but overdispersion is present in both cases
and follows a clear sample-wide trend.
See the \emph{dispersion survey} section of
Protocol~S\ref{supp:rmdzip}
for many more examples of overdispersed microbiome data
than the three included in Figure~\ref{fig:overdispersion}.
The consistent (though non-linear) relationship between variance and mean
indicates that parameters of a NB model, especially $\phi_{i}$,
can be adequately estimated among biological replicates
of microbiome data,
despite a previous weak assertion to the contrary~\cite{metastats}.


In simulations evaluating clustering accuracy, 
we found that rarefying undermined the performance
of downstream clustering methods.
This was the result of omitted read counts,
added noise from the random sampling step in rarefying, 
as well as omitted samples with small library sizes
that nevertheless were accurately clustered
by alternative procedures on the same simulated data
(Figure~\ref{fig:clustaccuracy}).
The extent to which the rarefying procedure performed worse
depended on 
the effect-size (ease of clustering), 
the typical library size of the samples in the simulation,
and the choice of threshold for the minimum library size
(Figure~\ref{fig:clustaccuracyrare}).
We also evaluated the performance of alternative
clustering methods,
k-means and hierarchical clustering,
on the same tasks and found similar overall results
(Protocol~S\ref{supp:rmdzip}).

In additional simulations we investigated 
the dependency of clustering performance
on the choice of minimum library threshold, $N_{L,min}$.
We found that samples were trivial to cluster
for the largest library sizes using most distance methods,
even with the threshold set to the smallest library in the simulation
(no samples discarded, all correctly clustered).
However, at more modest library sizes 
typical of highly-parallel experimental designs
the optimum choice of size threshold is less clear.
A small threshold implies retaining more samples
but with a smaller number of reads (less information) per sample;
whereas a larger threshold implies more discarded samples, 
but with larger libraries for the samples that remain.
In our simulations the optimum choice of threshold 
hovered around the 15$^{th}$-percentile of library sizes
for most simulations and normalization/distance procedures
(Figure~\ref{fig:clustaccuracyrare}),
but this value is not generalizable to other data.
Regions within Figure~\ref{fig:clustaccuracyrare}
in which all distances have converged 
to the same line ($y=1-x$)
are regions for which the minimum library threshold
completely controls clustering accuracy
(all samples not discarded are accurately clustered).
Regions to the left of this convergence 
indicate a compromise between discarding fewer samples
and retaining enough counts per sample for accurate clustering.

\begin{figure*}[!htbp]
\begin{center}
\includegraphics[width=6in]{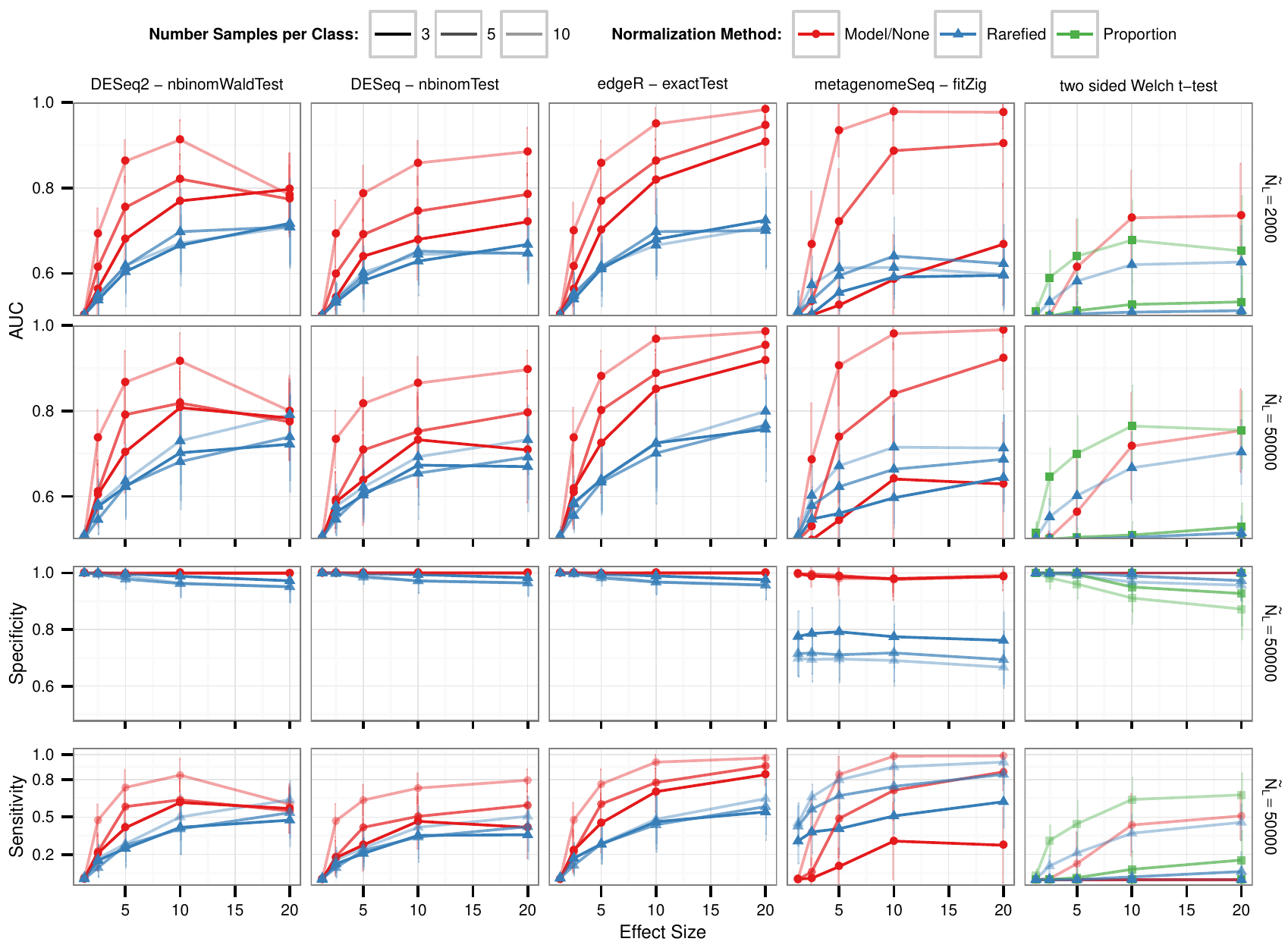}
\end{center}
\caption{
{\bf Performance of differential abundance detection with and without rarefying.} 
Performance summarized here by the ``Area Under the Curve'' (AUC) metric 
of a Receiver Operator Curve (ROC)~\cite{ROCR} (vertical axis).
Briefly, the AUC value varies from 0.5 (random) to 1.0 (perfect),
and incorporates both sensitivity and specificity.
The horizontal axis indicates the effect size, 
shown as the actual multiplication factor
applied to the OTU abundances. 
Each curve traces the respective normalization method's mean performance of that panel,
with a vertical bar indicating a standard deviation in performance 
across all replicates and microbiome templates. 
The right-hand side of the panel rows indicates
the median library size, $\tilde{N}_L$,
while the darkness of line shading indicates the number of samples per simulated experiment.
Color shade and shape indicate the normalization method. 
See Methods section for the definitions of each normalization and testing method.
All P-values were adjusted for multiple hypotheses using BH~\cite{BenjaminiHochberg},
and a detection significance threshold of 0.05.
}
\label{fig:diffabundauc}
\end{figure*}

In simulations evaluating performance in the detection of differential abundance,
we found an improvement
in sensitivity and specificity 
when normalization and subsequent tests are based upon
a relevant mixture model
(Figure~\ref{fig:diffabundauc}).
Multiple t-tests with correction for multiple inference
did not perform well on this data,
whether on rarefied counts or on proportions.
A direct comparison of the performance
of more sophisticated parametric methods
applied to both original and rarefied counts
demonstrates the strong potential of these methods
and large improvements in sensitivity and specificity
if rarefying is not used at all.

In general, the rate of false positives
from tests based on
proportions or rarefied counts 
was unacceptably high, 
and increased with the effect size.
This is an undesirable phenomenon in which
the increased relative abundance of the true-positive OTUs
(the effect)
is large enough that the null (unmodified) OTUs
appear significantly more abundant in the null samples 
than in the test samples.
This explanation is easily verified by 
the sign of the test statistics of the false positive OTU abundances,
which was uniformly positive
(Protocol~S\ref{supp:rmdzip}).
Importantly, this \emph{side-effect}
of a strong differential abundance 
was observed rarely in edgeR performance results
under TMM normalization (not shown) but not with RLE normalization (shown),
and was similarly absent in DESeq(2) results.
The false positive rate for edgeR and DESeq(2)
was near zero under most conditions,
with no obvious correlation
between false positive rate and effect size.
In most simulations count proportions outperformed rarefied counts
due to better sensitivity, 
but also suffered from a higher rate of false positives 
at larger values of effect size
(Figure~\ref{fig:diffabundauc}, Protocol~S\ref{supp:rmdzip}).


The rarefying normalization procedure was associated with performance costs 
in both sample-clustering and differential abundance statistical evaluations,
enumerated in the following.

\begin{enumerate}
  \item Rarefied counts represent only 
  	a small fraction of the original data,
	implying an increase in Type-II error --
	often referred to as \emph{a loss of power}
	or \emph{decreased sensitivity}
	(Table~\ref{tab:rarefyminimal}).
	In sample-wise comparisons, this lost power is evident
	through two separate phenomena, 
	(1) samples that cannot be classified because they were discarded,
	(2) samples that are poorly distinguishable
	because of the discarded fraction of the original library
	(Figure~\ref{fig:clustaccuracyrare}).
	Differential abundance analyses that include
	moderate to rare OTUs are even more sensitive 
	to this loss of power, 
	where rarefied counts perform worse
	in every analysis method we attempted
	(Figure~\ref{fig:diffabundauc}, Protocol~S\ref{supp:rmdzip}).
  \item Rarefied counts remain overdispersed
  	relative to a Poisson model,
	implying an increase in \mbox{Type-I} error (decreased specificity).
	Overdispersion is theoretically expected for counts of this nature,
	and we unambiguously detected overdispersion in our survey
	of publicly available microbiome counts (Figure~\ref{fig:overdispersion}).
	Estimating overdispersion is also more difficult after rarefying
	because of the lost information
	(Figure~\ref{fig:diffabundauc}).
	In our simulations, Type-I error was much worse 
	for rarefied counts than original counts
	(Figure~\ref{fig:diffabundauc}, Protocol~S\ref{supp:rmdzip}).
  \item Rarefying counts requires an arbitrary selection
	of a library size minimum threshold
	that affects downstream inference 
	(Figure~\ref{fig:clustaccuracyrare}),
	but for which an optimal value 
	cannot be known for new empirical data~\cite{NavasMolina:2013bc}.
  \item The random aspect of subsampling is unnecessary 
  	and adds artificial uncertainty
	(Protocol~S\ref{supp:rmdzip}, 
	minimal example, bottom).
	A superior transformation (though still inadmissible)
	is to instead round the expected 
	value of each count at the new smaller library size,
	that is $\parallel{}K_{ij}N_{L,min}/s_{j}\parallel$,
	avoiding the additional sampling error 
	as well as the need to
	repeat the random step~\cite{Lozupone:2010dh}
	and publish the random seed/process.
\end{enumerate}

Due to these demonstrated limitations and proven sub-optimality, 
we advocate that 
\textbf{rarefying should not be used}.
In special cases the costs listed above may be acceptable
for sample-comparison experiments 
in which the effect-size(s) and the original library sizes 
are large enough to withstand the loss of data.
Many early descriptive studies fall into this category --
for example comparing functionally distinct 
human body sites or environments~\cite{globalpatterns} --
and the ability to accurately distinguish
those vastly-different microbiome samples is not in question,
even with rarefied counts.
However, for new empirical data the effect size(s) are unknown
and may be subtle;
and consequently, rarefying may undermine downstream analyses.

In the case of differential abundance detection,
it seems unlikely that the cost of rarefying is ever acceptable.
In our simulations, both rarefied counts and sample proportions resulted in
an unacceptably high rate of false positive OTUs.
As we described theoretically in the introduction,
this is explained by differences among biological replicates 
that manifest as overdispersion, 
leading to a subsequent underestimate of the true variance
if a relevant mixture model is not used.
We detected overdispersion among biological replicates
in all publicly available microbiome count datasets that we surveyed
(Figure~\ref{fig:overdispersion}, Protocol~S\ref{supp:rmdzip}).
Failure to account for this overdispersion -- 
by using proportions or rarefied counts --
results in a systematic bias 
that increases the Type-I error rate
even after correcting for multiple-hypotheses
(\emph{e.g.} Benjamini-Hochberg~\cite{BenjaminiHochberg}).
In other words, 
if overdispersion has not been addressed,
we predict many of the reported differentially abundant OTUs
are false positives attributable to an underestimate of uncertainty.

In our simulations 
this  propensity for Type-I error
\emph{increased} with the effect size,
\emph{e.g.} the fold-change in OTU abundance 
among the true-positive OTUs.
For rarefied counts, 
we also detected a simultaneous increase in Type-II error
attributable to the forfeited data.
It may be tempting to imagine that 
the increased variance estimate due to rarefying
could be counterbalanced by the variance underestimate 
that results from omitting a relevant mixture model.
However, such a scenario constitutes an unlikely special case,
and false positives will not compensate for the false negatives in general.
In our simulations both Type-I and Type-II error 
increased for rarefied counts
(Figure~\ref{fig:diffabundauc}, Protocol~S\ref{supp:rmdzip}).

Fortunately, we have demonstrated
that strongly-performing alternative methods 
for normalization and inference are already available.
In particular, an analysis that models counts with the Negative Binomial --
as implemented in DESeq2~\cite{DESeq} or in edgeR~\cite{edgeR} with RLE normalization --
was able to accurately and specifically detect differential abundance
over the full range of
effect sizes, 
replicate numbers,
and library sizes that we simulated
(Figure~\ref{fig:diffabundauc}).
DESeq-based analyses are routinely applied 
to more complex tests and experimental designs
using the generalized linear model interface in R~\cite{Rglm},
and so are not limited to a simple two-class design. 
We also verified an improvement in differential abundance performance 
over rarefied counts or proportions 
by using an alternative mixture model based on the zero-inflated Gaussian,
as implemented in the metagenomeSeq package~\cite{metagenomeSeq}.
However, we did not find that metagenomeSeq's AUC values were uniformly highest,
as Negative Binomial methods had higher AUC values
when biological replicate samples were low.
Furthermore, while metagenomeSeq's AUC values 
were marginally higher than Negative Binomial methods
at larger numbers of biological replicates,
this was generally accompanied
with a much higher rate of false positives
(Figure~\ref{fig:diffabundauc}, Protocol~S\ref{supp:rmdzip}).

Based on our simulation results
and the widely enjoyed success for highly similar RNA-Seq data,
we recommend using DESeq2 or edgeR to perform 
analysis of differential abundance in microbiome experiments.
It should be noted that we did not comprehensively explore
all available RNA-Seq analysis methods,
which is an active area of research.
Comparisons of many of these methods 
on empirical~\cite{Nookaew:2012co,Bullard2010} 
and simulated~\cite{TCC, sSeq, SAMSeq} data 
find consistently effective performance 
for detection of differential expression.
One minor exception is an increased Type-I error for edgeR
compared to later methods~\cite{Nookaew:2012co},
which was also detected in our results
relative to DESeq and DESeq2 when TMM normalization was used
(not shown)
but not after switching to RLE normalization
(Figure~\ref{fig:diffabundauc}, Protocol~S\ref{supp:rmdzip}).
Generally speaking, 
the reported performance improvements between these methods
are incremental relative to the large gains
attributable to applying a relevant mixture model of the noise
with shared-strength across OTUs (features).
However, some of these alternatives from the RNA-Seq community
may outperform DESeq on microbiome data meeting special conditions,
for example a large proportion of true positives and sufficient replicates~\cite{baySeq},
small sample sizes~\cite{sSeq},
or extreme values~\cite{DFI}.

Although we did not explore the topic
in the simulations here described,
a procedure for further improving 
differential expression detection performance,
called \emph{Independent Filtering}~\cite{Bourgon:2010cr},
also applies to microbial differential abundance.
Some heuristics for filtering low-abundance OTUs
are already described in the documentation of 
various microbiome analysis workflows~\cite{mothur,caporaso2010qiime},
and in many cases these can be classified
as forms of Independent Filtering.
More effort is needed to optimize Independent Filtering
for differential abundance detection,
and rigorously define the theoretical basis and heuristics
applicable to microbiome data.
Ideally a formal application of Independent Filtering of OTUs
would replace many of the current \emph{ad hoc} approaches
that often includes poor reproducibility and justification,
as well as the opportunity to introduce bias.


Some of the justification for the rarefying procedure 
has originated from exploratory sample-wise comparisons of microbiomes 
for which it was observed that a larger library size also results
in additional observations of rare species, 
leading to a library size dependent increase 
in both alpha-diversity measures and beta-diversity dissimilarities~\cite{Chao:2005ue,Lozupone:2010dh}, 
especially UniFrac~\cite{Schloss:2008hd}. 
It should be emphasized that this represents a failure 
of the implementation of these methods to properly account for rare species
and not evidence that diversity depends on library size.
Rarefying is far from the optimal method for addressing rare species, 
even when analysis is restricted solely to sample-wise comparisons.
As we demonstrate here, it is more data-efficient
to model the noise and address extra species
using statistical normalization methods
based on variance stabilization and robustification/filtering.
Though beyond the scope of this work,
a Bayesian approach to species abundance estimation
would allow the inclusion of pseudo-counts from a Dirichlet prior
that should also substantially decrease this sensitivity.

Our results have substantial implications 
for past and future microbiome analyses,
particularly regarding the interpretation of differential abundance.
Most microbiome studies 
utilizing high-throughput DNA sequencing 
to acquire culture-independent counts of species/OTUs
have used either proportions or rarefied counts 
to address widely varying library sizes. 
Left alone, both of these approaches suffer from a failure
to address overdispersion among biological replicates, 
with rarefied counts also suffering from a loss of power,
and proportions failing to account for heteroscedasticity.
Previous reports of differential abundance based on rarefied counts or proportions
bear a strong risk of bias toward false positives,
and may warrant re-evaluation. 
Current and future investigations
into microbial differential abundance
should instead model uncertainty using a hierarchical mixture,
such as the Poisson-Gamma or Binomial-Beta models,
and normalization should be done using the
relevant variance-stabilizing transformations.
This can easily be put into practice
using powerful implementations in R, 
like DESeq2 and edgeR,
that performed well on our simulated microbiome data.
We have provided wrappers for edgeR, DESeq, DESeq2, and metagenomeSeq
that are tailored for microbiome count data
and can take common microbiome file formats
through the relevant interfaces in the phyloseq package~\cite{phyloseqplosone}.
These wrappers are included with 
the complete code and documentation
necessary to exactly reproduce
the simulations, analyses, surveys, and examples shown here,
including all figures
(Supplementary Information File~S\ref{supp:rmdzip}).
This example of fully reproducible research
can and should be applied to future publication 
of microbiome analyses~\cite{Gentleman:2004vk,Peng02122011,Donoho:2010cx}.

\section*{Acknowledgments}

We would like to thank the developers of the open source packages
leveraged here for improved insights into microbiome data,
in particular Gordon Smyth and his group
for edgeR~\cite{edgeR}, to Mihail Pop and his team for
metagenomeSeq~\cite{metagenomeSeq} 
and Wolfgang Huber and his team for DESeq and DESeq2~\cite{DESeq};
whose useful documentation and continued support have been invaluable.
The Bioconductor and \R{} teams~\cite{Bioconductor,Rlanguage}
have provided valuable support for our development and release
of code related to microbiome analysis in R.
We would also like to thank Rob Knight and his lab for QIIME~\cite{caporaso2010qiime},
which has drastically decreased the time required to get
from raw phylogenetic sequence data to OTU counts.
Hadley Wickham created and continues to support
the ggplot2~\cite{ggplot2} and reshape~\cite{reshape}/plyr~\cite{plyr} packages
that have proven useful for graphical representation and manipulation of data, respectively. 
RStudio and GitHub have provided immensely useful and free applications
that were used in the respective development and versioning 
of the source code published with this manuscript.

\clearpage
\fancyhead[R]{References}
\bibliography{norarefy-revised}

\onecolumn 
\captionsetup{labelformat=empty, labelsep=none} 
\setcounter{figure}{0}

\clearpage
\fancyhead[L]{Text~S\ref{supp:mathstat}: Mathematical Supplement} 
\fancyhead[R]{2013 McMurdie and Holmes}
\setcounter{page}{1} 
\section*{Text~S\ref{supp:mathstat}: Mathematical Supplement}

\begin{figure}[!htbp]
\raggedleft
\captionsetup{font={Large},justification=justified}
\caption{
A supplemental appendix of statistical mathematics supporting this article.
}
\label{supp:mathstat}
\end{figure}

In this supplementary material we go over some
of the statistical details pertaining to the use 
of hierarchical mixture models
such as the Negative Binomial and the Beta Binomial,
which are appropriate for addressing
additional sources of variability
inherent to microbiome experimental data,
while still retaining statistical power.
We have concentrated our comparison efforts on the Gamma-Poisson
mixture model as some authors\cite{lu2005} have remarked that this approach
seems to be the most  statistically robust
approach in the sense that the presence of outliers and model misspecification
does not over-perturb the results.
We show how a Negative Binomial distribution can occur
in different ways leading to different parameterizations.
We then show that there are transformations we can apply to 
these random variables, such that the transformed data have a
variance which is much closer to constant than the original.
These \emph{variance stabilizing transformations}
lead to more efficient estimators 
and give better decision rules than those obtained
via the normalization-through-subsampling method known as \emph{rarefying}.

\subsection*{
Two parameterizations of the negative binomial}

In classical probability, the negative binomial is often introduced
as the distribution of the number of successes in a sequence of Bernoulli trials 
with probability of success $p$ before the number $r$ failures occur.
Thus with the two parameters $r$ and $p$, 
the probability distribution for the negative binomial is given as 

\begin{displaymath}  X \sim \operatorname{NB}(r;p) \end{displaymath}
\begin{eqnarray*}
  P(X=k)&=&{k+r-1 \choose k} (1-p)^r p^k \\
  &=&\frac{\Gamma(k+r)}{k! \Gamma(r)}(1-p)^r p^k
\end{eqnarray*}
The mean of the distribution is $m=\frac{pr}{1-r}$ and the variance
$Var(X)=\frac{pr}{(1-p)^2}$. Sometimes the distribution is given a different
parameterization which we use here.
This takes as the two parameters: the mean $m$ and
$r=\frac{1-p}{p} m$, then the probability mass distribution is
rewritten:
\begin{displaymath}X \sim \operatorname{NB}(m;r) \end{displaymath}
\begin{eqnarray*}
P(X=k)&=&{k+r-1 \choose k} \left(\frac{r}{r+m}\right)^r \left(\frac{m}{r+m}\right)^k\\
  &=&\frac{\Gamma(k+r)}{k! \Gamma(r)}
  \left(\frac{r}{r+m}\right)^r \left(\frac{m}{r+m}\right)^k
\end{eqnarray*}
The variance is  $Var(X)=\frac{m(m+r)}{r}=m+\frac{m^2}{r}$,
we will also use $\phi=\frac{1}{r}$ and call this the overdispersion parameter, giving
$Var(X)=m+\phi m^2$.
When $\phi=0$ the distribution of $X$ will be Poisson(m). 
This is the (mean=m,overdispersion=$\phi$) parametrization we will use from now on.

\subsection*{Negative Binomial as a
hierarchical mixture for read counts}

In biological contexts such as RNA-seq and microbial count data the
negative binomial distribution arises as a hierarchical 
mixture of Poisson distributions.
This is due to the fact that if we had technical replicates with the same
read counts, we would see Poisson variation with a given mean.
However, the variation among biological replicates
and library size differences
both introduce additional sources of variability.

To address this, we take the means of the Poisson variables 
to be random variables themselves having a Gamma distribution 
with (hyper)parameters shape $r$ 
and scale $p/(1-p)$.
We first generate a random mean, $\lambda$, for the Poisson
from the Gamma, and then a random variable, $k$, from the Poisson($\lambda$).
The marginal distribution is:
\begin{eqnarray*}
P(X=k) &=&
\int_0^\infty Po_\lambda(k) \times \gamma_{(r,\frac{p}{1-p})} d\lambda\\
&=&\int_0^\infty \frac{\lambda^k}{k!}e^{-\lambda} \times
\frac{\lambda^{r-1}e^{-\lambda\frac{1-p}{p}}}{(\frac{p}{1-p})^r \Gamma(r)}
d\lambda \\
&=& \frac{(1-p)^r}{p^rk!\Gamma(r)} \int_0^\infty \lambda^{r+k-1}e^{-\lambda/p}
d\lambda \\
&=& \frac{(1-p)^r}{p^rk!\Gamma(r)} p^{r+k}\Gamma(r+k)\\
&=& \frac{\Gamma(r+k)}{k!\Gamma(r)} p^{k} (1-p)^r
\end{eqnarray*}

\subsection*{Variance Stabilization}
Statisticians usually prefer to deal with errors across samples or in regression
situations which are independent and identically distributed. In particular
there is a strong preference for homoscedasticity (equal variances) across
all the noise levels. This is not the case when we have unequal sample
sizes and variations in the accuracy across instruments. A standard way of dealing
with heteroscedastic noise is to try to decompose the sources of heterogeneity
and apply transformations that make the noise variance almost constant.
These are called \emph{variance stabilizing transformations}.

Take for instance different Poisson variables with mean $\mu_i$.
Their variances are all different if the $\mu_i$ are different.
However, if the square root transformation is applied to each of the variables,
then the transformed variables will have approximately constant variance\footnote{Actually if we take the transformation $x \longrightarrow 2\sqrt{x}$ we obtain a variance approximately equal to 1.}.
More generally, choosing a transformation that makes the variance constant
is done by using a Taylor series expansion, called the delta method.
We will not give the complete development of variance stabilization in the context
of mixtures but point the interested reader to 
the standard texts in Theoretical statistics such as
\cite{Rice2007} and one  of the original articles on variance stabilization\cite{Anscombe1948}.
Anscombe showed that there are several transformations that stabilize the variance
of the Negative Binomial depending on the values of the parameters $m$ and $r$,
where $r=\frac{1}{\phi}$, sometimes called the \emph{exponent} of the Negative Binomial.
For large $m$ and constant $m\phi$, the transformation
\begin{displaymath}
\sinh^{-1}
\sqrt{\left(\frac{1}{\phi}-\frac{1}{2}\right)\frac{x+\frac{3}{8}}{\frac{1}{\phi}-\frac{3}{4}}}
\end{displaymath}
gives a constant variance around $\frac{1}{4}$.
Whereas for $m$ large and $\frac{1}{\phi}$ not substantially increasing,
the following simpler transformation is preferable
\begin{displaymath}
\log\left(x+\frac{1}{2\phi}\right)
\end{displaymath}

These two transformations are actually used in what is often known as
a \emph{generalized logarithmic} transformation
applied in microarray variance stabilizing transformations and
RNA-seq normalization\cite{DESeq}.
\subsection*{Modeling read counts}
If we have technical replicates with the same number of reads $s_j$, we expect
to see  Poisson variation with mean $\mu=s_j u_i$,
for each taxa $i$ whose
 incidence proportion we denote by
$u_i$. 
Thus the number of reads for the sample $j$ and taxa $i$ would be
\begin{displaymath}K_{ij} \sim \mbox{ Poisson }(s_{j}u_i)\end{displaymath}
We use the notational convention that lower case letters designate
fixed or observed values whereas upper case letters designate random variables.

For biological replicates within the same group --
such as treatment or control groups or the same environments --
the proportions $u_i$ will be variable between samples.
A flexible model that works well for this variability is
the Gamma distribution, as it has two parameters
and can be adapted to many distributional shapes.
Call the two parameters $r_i$ and $\frac{p_i}{1-p_i}$. So that
$U_{ij}$ the proportion of taxa $i$ in sample $j$
is distributed according to Gamma$(r_i,\frac{p_i}{1-p_i})$.
Thus we obtain that the read counts $K_{ij}$ have a Poisson-Gamma
mixture of different Poisson variables. 
As shown above we can use the Negative Binomial
with parameters
$(m=u_{i}s_j)$ and $\phi_{i}$ as a satisfactory model of the variability.

Now we can add to this model the fact that the samples belong to different conditions
such as treatment and control or different environments.
This is done by separately estimating the values of the parameters,
for each of the different biological replicate conditions/classes.
We will use the index $c$ for the different conditions,
we then have the counts for the taxa $i$ and sample $j$ in condition $c$
having a Negative Binomial distribution
with $m_{c}=u_{ic}s_j$
and $\phi_{ic}$ so that the variance is written

\begin{equation} \label{eq:variance}
u_{ic}s_j + \phi_{ic} s_j^2 u_{ic}^2
\end{equation}

We can estimate the parameters $u_{ic}$ and $\phi_{ic}$ from the data
for each OTU and sample condition.
This is usually best accomplished by leveraging information across OTUs --
taking advantage of a systematic relationship between 
the observed variance and mean --
to obtain high quality shrunken estimates.
The end result provides a variance stabilizing transformation of the data
that allows a statistically efficient comparisons between conditions.
This application of a hierarchical mixture model
is very similar to the random effects models used in the context of
analysis of variance.
A very complete comparison of this particular choice of Gamma-Poisson mixture to the Beta-Binomial
and nonparametric approaches can be found in
\cite{sSeq}.

By comparison, the procedures involving a systematic downsampling (rarefying)
are inadmissible in the statistical sense,
because there is another procedure that dominates it
using a mean squared error loss function.
With a Bayesian formalism we can show
that the hierarchical Bayes model gives a Bayes rule
that is admissible\cite{Berger}.

\subsection*{Other mixture models}
If instead of modeling the read counts 
one uses the proportions as the random variables, 
with differing variances due to different library sizes,
the Beta-Binomial model is the standard approach. 
This has also been used for RNA-seq data \cite{BBSeq}
and the package metaStats\cite{metastats} 
uses this model although they don't
use variance stabilizing transformations of the data.

\clearpage
\fancyhead[L]{Protocol~S\ref{supp:rmdzip}: Source Code} 
\begin{figure}[!htbp]
\begin{center}
\end{center}
\caption{
{\bf Protocol~S\ref{supp:rmdzip}.
A zip file containing all supplementary source files.}
This includes the Rmd source code, HTML output, and all related documentation and code to completely and exactly recreate every results figure in this article. 
}
\label{supp:rmdzip}
\end{figure}
%
%
%



\end{document}